# Applications of fMRI for Brain Mapping

Nivedita Daimiwal[1,2]
[1]Research Scholar, Sathyabama University, Chennai, INDIA
[2]Cummins college of Engg. For Women, Pune
nivedita.daimiwal@gmail.com

Dr. M. Sundhararajan
Principal, Shri Laxmi Ammal Engineering College, Chennai, INDIA
msrajan69@gmail.com

Revati Shriram
Cummins College of Engg. For Women, Pune, INDIA
revatishriram@yahoo.com

*Abstract—* Brain-mapping techniques have proven to be vital in understanding the molecular, cellular, and functional mechanisms of the brain. Normal anatomical imaging can provide structural information on certain abnormalities in the brain. However there are many neurological disorders for which only structure studies are not sufficient. In such cases it is required to investigate the functional organization of the brain. Further it is necessary to study the brain functions under normal as well as diseased conditions. Brain mapping techniques can help in deriving useful and important information on these issues. Brain functions and brain area responsible for the particular activities like motor, sensory speech and memory process could be investigated. The authors provide an overview of various Brain Mapping techniques and fMRI signal processing methods.

## I. INTRODUCTION

Modern imaging has transformed practice in the clinical neurosciences by providing information about structural abnormalities in the brain non invasively. However many chronic neurological or psychiatric complaints confronted in the clinic (for example pain, movement, disorders, depression and psychosis) are not associated with structural abnormalities that can be detected in an individual patient with current clinical technologies. There are many approaches to measurement of functional changes in brain. While some method directly monitor electrical events in neurons, others by secondary effects of increased neuronal firing rates. Metabolic demand and the requisite changes in blood delivery are useful for localizing the sites and magnitude of brain activity. Several functional brain mapping techniques have been developed over the 3 decades which have revolutionized our ability to map activity in the living brain. [1, 22]

## II. BRAIN MAPPING METHODS

A. Functional MRI (fMRI): Is one of the techniques that is used to identify the brain regions that are associated with certain motor or sensory tasks. The most common fMRI techniques used to capture functional images of the brain employs Blood Oxygenation level Dependent (BOLD) contrast. In the BOLD effect, a neural activity in the brain caused by some sensory or motor tasks produces localized changes in the blood flow and hence the resulting oxygenation level is subjected to variations. Whenever some task is performed, neuronal in these areas also increases followed by an increase in glucose and oxygen rates. The hemodynamic and metabolic changes associated with brain functions affect the deoxyhaemoglobin contents in the tissue. This gives rise to a contrast that can be detected using the MRI scanner. Functional activation in fMRI studies by mapping changes in cerebral venous oxygen concentration that correlate with neuronal activity. Such approaches require fast two dimensional brain imaging. Using echo planar imaging two dimensional slice data can be acquired in 40 ms with an in plane anatomical resolution of about 1 mm. thus functional maps of the human brain can be obtained without ionizing radiation or the administration of exogenous contrast material.[2]

B. Positron Emission Tomography (PET): RP (Radio pharmaceuticals) labeled with positron emitting radionuclides are used. Positron emitting radionuclides like C-11, N-13, O-15 & F-18 are used in PET RP, as all these are bio-molecules (F-18 mimics like Hydrogen). One of the stable bio-molecule can be replaced by positron emitting one to study the in-vivo biochemistry. PET plays vital role in Oncology, Cardiology and Neurology. Since all these radioisotopes are cyclotron produced and short lived it is essential to have the cyclotron in the vicinity. Table 1 shows the RP used in PET.[4]

| RP USED IN PET | | |
|---|---|---|
| RPT1/2(min) | Energy Max (KeV) | Range(mm) |
| F-18   110 | 640 | 2.39 |
| C-11   20 | 960 | 4.108 |
| N-13   10 | 1190 | 5.39 |
| O-15   2 | 1720 | 8.2 |

Table 1: RP used in PET

Two 511 KeV photons, ejected at $180^\circ$ apart from annihilation of positron, are used for tomographic imaging. Two scintillation detectors which are connected in coincidence circuit are placed $180^\circ$ apart. The event simultaneously detected (with in 6-12 nSec) by these detector are considered as 'True' events. If two photons each generated from different annihilation process interacts with two detectors in coincidence simultaneously then they are called as ''Random'. Scattered and Random events are unwanted since they degrade the quality of image. PET is used to measure cerebral metabolism, blood flow and volume, oxygen utilization, neurotransmitter synthesis, and receptor binding. The spatial resolution of PET is approximately 5 mm/voxel. [2]

B. Single Photon Emission Tomography (SPECT): SPECT uses radiopharmaceuticals administered intravenously or by inhalation to evaluate function in human brain. These radiopharmaceuticals incorporate isotopes including xenon -133, iodine-123, technesium-99m, and others that emit single photon radiation, most typically in the form of gamma rays. SPECT techniques have a current resolution of approximately 9mm/voxel.[2]

C. Electroencephalography (EEG): The signals reflect signals reflect neuronal activity in the superficial layer of the cerebral cortex and the accompanying distortion by volume conductance within tissue and through the skull. Spatial resolution of the technique is determined by the density of electrode placements but typically is on the order of a few square centimeters at the cortical surface. [2, 3]

D. Magnetoencephalography (MEG): Takes advantage of the fact that the weak electrical fields in the brain that are detected by EEG also induce a magnetic field that can be externally measured. The extremely low magnitude of these fields requires the use of supercooled devices in rooms that are isolated from the external magnetic and electrical environment. Since magnetic rather than electrical fields are detected using MEG, distortion caused by the effects of the skull are eliminated. EEG and MEG has a temporal resolution that is in the millisecond time frame. [2, 3]

E. Optical Intrinsic Imaging Techniques: Optical imaging of intrinsic signals maps the brain by measuring intrinsic activity related changes in tissue reflectance, functional physiological changes in tissue reflectance. Functional physiological changes, such as increases in blood volume, hemoglobin oxymetry changes, and light scattering changes, result in intrinsic tissue reflectance changes that are exploited to map functional brain activity. [2]

fMRI is a non invasive method for investigating the structure and function of the brain. It has good temporal and spatial resolutions and it is therefore possible to carry out an fMRI experiment in a repetitive manner. The sensitivity that is possible with fMRI is sufficient for detecting the transient changes in the deoxyhaemoglobin content.[1]

III. TYPICAL FMRI EXPERIMENT

An fMRI experiment can be performed on the same 1.5 T MR scanner that is used for routine work. A fast MR imaging technique such as Echo-planar imaging (EPI) [5] is employed in order to detect the neural activity and the resulting oxygenation levels. It is important here that a single image obviously does not give any functional information. In fact, it is the variation of the image intensity levels when recorded with respect to time that contains the desired functional information. Therefore in fMRI, a number of images of the brain are recorded consecutively with respect to time in a single fMRI experiment. This is shown in figure 1. [1]

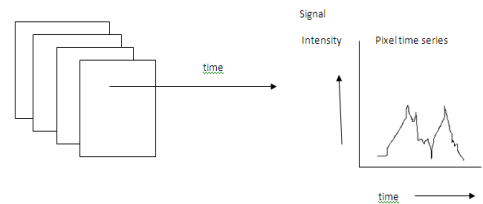

Figure 1: Sequence of Images Recorded with Respect to Time in fMRI and a Pixel Time Series [1]

fMRI is a relative technique in the sense that it compares the images taken during two different states of the task. During the ON state the subject performs some task (the activation state) where as no task is performed during OFF state (the base line state), several such cycles of activation and a baseline signal.

Essentially the signal is determined by the difference between the intensities of the images recorded during the ON state and the intensities recorded during the OFF state. Images recorded during the activation periods and those recorded during the baseline states are then compared. Typically a mean difference image is formed and then tests for statistical significance are carried out to obtain the activation maps. Activation maps show the brain regions that are responsible for a given sensory or motor task. This provides a meaningful picture of the neural activity from the perspective of the brain function.

### IV. FMRI SIGNAL PROCESSING METHODS

There are a number of signal processing methods useful in processing the fMRI data:

Bendettini et.al. [6] have suggested a method in 1993 that uses both time and frequency domain information also know as temporal spatial and spectral spatial representation. Temporal cross correlation function, Fourier analysis and thresholding techniques are used for data processing. After observing the cross correlation of the time series with the stimulus function, the correlation coefficient for each pixel is calculated and then mapped onto a correlation coefficient map.

In 1994, Friston et.al [7] described a linear model for the haemodynamic response present in an fMRI time series. Simple model of a linear time invariant system based on convolution. The problem was to select the haemodynamic response function so that the convolution of the stimulus function with the haemodynamic response would give the activation signal. Friston et.al [8][9] explained how to model and detect the activations in fMRI time series in 1995. Worsley and Friston described statistical parametric mapping (SPM) that uses a generalized linear model (GLM) operating at each voxel. The aim of the problem that was described in [7] was to estimate the parameter β of the linear model.

$$x = \beta G + e$$

Where x is the unsmoothed time series and e the error vector whose components are independent and normally distributed with zero mean and variance 1. This model consists of a design matrix that is common to all the voxels and set of parameter estimates that are voxel specific. The design matrix contains the information about the activation paradigm and the confounding variables. The data that is spatially smoothed using a Gaussian filter and the GLM are fitted to each voxel. Then a t-statistic is used for detecting the significantly activated pixels.

In 1996, Bullmore et.al. Investigated statistical methods for the estimation and inference of the fMRI data [10]. They have suggested that the haemodynamic response differs from location to location and that the haemodynamic response delay is spatially varying.

In 1997, K.J Worsley [11] described a new method based on multivariate linear models that could overcome the drawbacks of the methods that used Scaled Subprofile Models (SSM), Singular Value Decomposition (SVD), Partial Least Squares (PLS) And Canonical Variates Analysis. A model that incorporates a spatially varying haemodynamic response. Use of the Discrete Fourier Transform (DFT) of fMRI time series at each voxel was made for analyzing the fMRI data set. In 1998,

Ogawa et.al [12] discussed the various aspects of characterizing the functional MRI elaborating the relation between MRI signals and neural evens. Addressing the issue of activation signal detection, Ruttiman and Unser et.al [13] in 1998, used the Discrete Wavelet Transform of the mean difference image in the spatial domain and then applied statistical analysis for obtaining the activation map. Periodicity assumption of an fMRI time series model was proposed by Babak A. Ardekani and Iwao Kanno [14] and a truncated Fourier series was used in their model.

A periodic signal detection method for fMRI data was proposed by Lars Kai Hansen, and Jan Larsen[15]. Their method used a Bayesian framework to detect periodic components in the fMRI data. Assessment of fMRI activation signal detection in the wavelet domain with different wavelets was reported by M.Desco et.al [16]. A newer approach for improving the signal to noise ratio for fMRI data analysis was demonstrated by Muller et.al [17]. A hierarchical clustering analysis method was used to select a cluster of pixels to improve the signal to noise ratio.

Cluster analysis using spectral peak statistics for selecting and testing the significance of activated fMRI time series was reported in the literature by Jarmasz and Somorjai [18]. The application of periodic stimulus, power spectrum ranked independent component analysis of periodic fMRI paradigm was carried out by Mortiz et.al [19]. The fundamental frequency of the periodic stimulus was considered and hence it again resembles purely a sinusoidal based approach .Ranking of spatial ICA components by magnitude contribution at this frequency of the stimulus was used for the detection purpose. Figure 2 shows the activation map. Colour pixels indicate the activated regions in the brain.

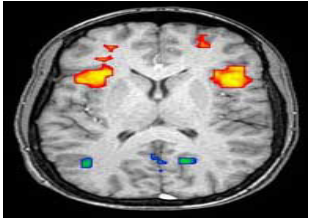

Figure 2: Activation map for one of the slices [20]

## V. CLINICAL APPLICATIONS

- Examine the anatomy of the brain.
- Determine precisely which part of the brain is handling critical functions such as thought, speech, movement and sensation, which is called brain mapping.
- Help assess the effects of stroke, trauma or degenerative disease (such as Alzheimer's) on brain function.
- Monitor the growth and function of brain tumors.
- Guide the planning of surgery, radiation therapy, or other surgical treatments for the brain. [20][2]

## VI. CONCLUSION

Neural activity and functional studies of the brain can be investigated with modalities like Positron Emission Tomography (PET), Single photon Emission Computed Tomography (SPECT), Magnetic Encephalography (MEG), and optical Imaging. All these modalities are compared briefly. fMRI is a non invasive method for investigating the structure and function of the brain. It has good temporal and spatial resolutions and it is therefore possible to carry out an fMRI experiment in a repetitive manner. fMRI based investigations, because of the safety, wide availability and extraordinary flexibility in terms of the application of this non ionizing imaging approach. Most recently, fMRI has emerged as a promising new extension of the technology for clinical neuroimaging.

ACKNOWLEDGMENT

The authors are grateful to Dr. Madhuri Khambete and Prof A. D. Gaikwad for their motivation, and help towards the completion of this paper, as well as for providing valuable advice.

We would also like to thank our colleagues from Instrumentation and Control Dept. of Cummins College of Engineering for Women for their feedback during the discussions.

REFERENCES

[1] "Functional Magnetic Resonance Imaging", novel transform method, Ajay V.Deshmukh, Vikram M.Gadre.
[2 ]"Brain Mapping the Methods", Arthur W. Toga, John C. Mazziotta
[3] Dale,A.M., and Sereno, M.I.(1993)."Improved localization of critical activity by combining EEG and MEG with MRI cortical surface reconstruction –A linear approach "J.cognit.Neurosci.5,162-176.
[4] Frackowiak, R.S., and Friston, K.J. (1994).Functional neuroanatomy of the human brain: Positron Emission tomography-A new neuroanatomical technique.J.Anat.184, 211-225.
[5] Michal K.stehling, Robert turner, Peter Mansfield, "Echo-planar Imaging Magnetic resonance imaging in a fraction of second", Science,vol.254,pp.44-49.
[6] P.A Bendettini, A.Jesmanowicz, E.C. Wong and J.S.Hyde," Processing strategies for time course data sets in functional MRI of the human brain", *Magn.Reson.Med. vol.30, pp.161-171, 1993.*
[7] K.J.Friston, P.Jezzard and R.Turner,"Analysis of functional time series", *Hum. Brain Mapp.Vol.1, pp.153-171, 1994.*
[8]K.J.Friston, A.P.Holmes,J.B.Polin,B.J.Grasby,C.R.Williams,R.S.J.Frackowiiak,nad R.Turner, "Analysis of time series Revisited", *Neuroimage,vol.2,pp.45-53,1995.*
**[9]** K.J.Friston, A.P.Holmes, K.J.Worsley, J.B.pline, C.D.frith, and R.S.J.Frackowiak,"Statistical Parametric Maps In Functional Imaging: A Generalized Linear Approach". Human Brain Map. Vol.2.pp.189-210,1995.
[10]E.T.Bullmore,M.J.Brammer,S.C.R.williams,S.Rabe-Henketh,N.Janot,A.david,et.al,"statistical Methods of estimation and Inference for Functional MR Image Analysis",Magn.Reson.Med.,vol.35,pp 261-277,1996.
[11] K.J.Worsley, J.B.Poline, K.J.Friston, and A.C.Evans," Characterizing the response of PET and fMRI data using multivariate linear models", Neuroimage,vol.6,pp.305-319,1997.
[12] S.Ogawa, R.S.Menon, S.G.Kim, and K. Ugurbil, "On the Characterizing o Functional Magnetic Resonance Imaging of the Brain",Annu.Rev.Biomol.Struct.,vol.27,pp.447-474,1998.
[13]Urs.E.Ruttimann, Michael Unser, Robert R.Rawlings, Daniel Rio, nick R.Ramsey, Venkata S.Mattay, Dniel W.Hommer, Oseph A.frank,and Daniel R.Wienberger, "Statisitical Analysis of functional MRI data in the Wavelet domin",*IEEE Trans.Med.Imag.*,vol.17,No2,pp.142-153,1998
[14]Babak A.Ardekani, and Iwao Kanno," Statistical methods for detection activation regions in functional MRI brain", M*ag.Reson.Imag.*, Vol 16,No10,pp.1217-1255,1998.

[15] Lars Kai Hansen, Nielsen and Jan Larsen,"Exploring fMRI Data for periodic Signal Components", AI *in medicine*, Vol25,no1,pp.35-44,2002

[16] M.Desco, J.A.Hernandez, A.Santosh and M.Brammer,"Multiresolution analysis in fMRI: "Sensitivity and Specificity in the Detection of Brain Activity", *Human Brain Map.*, vol.14,pp.16-27,2001.

[17] H.P.Muller, E.Kraft A.Lodolph, S.N.Erne, "New methods in fMRI Analysis" IEEE Engg.Med. and Bio., pp134-142,2002

[18] M.Jarmasz, R.L. Somorjai," Exploring regions of interest with clusters analysis using a spectral peak statistic for selecting and testing the significance of fMRI activation time series"' AI in Medicine,VOL.25,45-67,2002

[19] Chad H.Moritz, Baxter P.Rogers, and M.Elizabeth Meyerand,"Power spectrum ranked independent component analysis of a periodic fMRI complex Motor paradigm", Human Brain Mapping,vol.18,pp.111-122,2003.

[20] "Functional Brain Mapping", E. Mark Haacke, Ph.D.Washington University, St. Louis

[21] Brain-Mapping Techniques for Evaluating Post stroke Recovery and Rehabilitation: A Review

[22] "Applications of fMRI in translational medicine and clinical practice", Paul M.Matthew, Garry D.Honey and Edward T.Bullmore, September 2006, volume7.

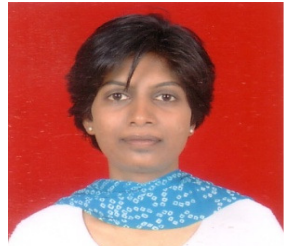

**Revati Shriram** received the BE Degree in Instrumentation and Control from University of Pune, MS in Electrical Engineering from Rose-Hulman Institute of Technology, Indiana, USA and she is currently working towards the PhD from Sathyabama University, Chennai.

She is currently working as an Assistant Professor in MKSSS's Cummins College of Engineering for Women, Pune, INDIA

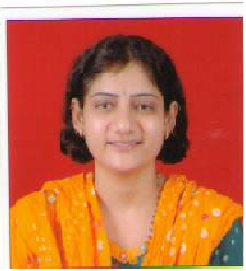

**Nivedita Daimiwal** received the B.E and M.E Degree in Biomedical Instrumentation from Pune University. She is currently working towards the PhD from Sathyabama University, Chennai. She is currently working as an Assistant Professor in MKSSS's Cummins College of Engineering for women, Pune, India.

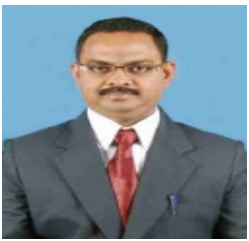

**Dr.M.Sundararanjan** received the M.S Degree from Birla Institute of Technology & Science (BITS), Pilani and PhD from Bharathidasan University, Trichy, INDIA. He is Currently Principal in Shri Laxmi Ammal College of Enginnering, Chennai.